\documentclass[12pt]{article}
\usepackage{amsmath}
\usepackage{breqn}
\usepackage{cite}
\usepackage{color}
\usepackage{graphicx}
\usepackage{amssymb}
\usepackage{adjustbox} 

\begin{document}

\title{\bf Friedmann equations from GUP-modified equipartition law}

\author{\"{O}zg\"{u}r \"{O}kc\"{u}  \thanks{Email: ozgur.okcu@yeniyuzyil.edu.tr}\\
{\small {\em Department of Electrical and Electronic Engineering,}}\\{\small {\em  Istanbul Yeni Yüzyıl University, TR34010 Istanbul, Türkiye}}}
\date{}
\maketitle
\begin{abstract}
In this paper, combining the thermodynamical arguments of the horizon with the quadratic generalised uncertainty principle (GUP), we heuristically obtain the modified equipartition law of energy. Employing this modified equipartition law of energy, we derive the Friedmann equations in Verlinde's entropic gravity. We find a maximum energy density at the beginning of the Universe. Remarkably, this feature emerges not only  for positive GUP parameter but also  for negative GUP parameter. From the initial acceleration, we deduce that the negative GUP parameter is more preferable. We also obtain maximum Hubble parameter from the first Friedmann equation,  indicating a universe without initial singularity. Moreover, we compute the Kretschmann curvature scalar, again indicating a non-singular universe. Interestingly, we find that GUP-modified Friedmann equations share some similarities with braneworld cosmolgy where the quadratic term in energy density appears. We also compute the deceleration parameter. Finally, we revisit the gravitational baryogenesis and show that the GUP-modified equipartition law of energy provides a mechanism for generating baryon asymmetry. Moreover, we constrain the GUP parameter from observations. 
\end{abstract}

\section{Introduction}
\label{introduction}

Based on the notion of black hole thermodynamics \cite{Bekenstein1972,Bekenstein1973,Bardeen1973,Hawking1974,Bekenstein1974,Hawking1975}, the gravity-thermodynamics conjecture has been a captivating research field since the seminal work of Jacobson \cite{Jacobson1995}. Considering the entropy-area relation $S=(k_{B}c^{3}A)/(4G_{N}\hbar)$ together with the Clausius relation $\delta Q=TdS$, he obtained the Einstein field equations \footnote{Throughout the paper, the physical constants such as $c$ (speed of light), $\hbar$ (reduced Planck constant), $G_{N}$ (Newton's gravitational constant) and $k_{B}$ (Boltzmann constant) are mostly kept explicit in order to make derivations more transparent and clarify dimensional consistency. In figures and tables, all physical constants set to unity for simplicity. In Section \ref{baryogeSect}, we use the units $\hbar=c=k_{B}=1$, following the Refs. \cite{Das2022BG,Feng2022,Luo2023}. Therefore, our results in Section \ref{baryogeSect} can be directly compared the results from the previous studies.}. Here, $\delta Q$ and $T$ are the energy flux and the Unruh temperature, respectively. The Einstein field equations are interpreted as an equation of state because this perspective treats the field equations as an entity from thermodynamic viewpoint. Since then, thermodynamical aspect of gravity has drawn much attention in the field and has been extensively studied in the literature \cite{Padmanabhan2002,Eling2006,Paranjape2006,Kothawala2007,Padmanabhan2007,Cai2005,Akbar2006,Akbar2007,Cai2007a,Cai2007b,Cai2008,Sheykhi2010a,Awad2014,Salah2017,Kouwn2018,Okcu2020,Okcu2024,Alsabbagh2023,Luo2023,Das2022BG,Feng2022,Luciano2025,Luciano2021,Luo2025,Sheykhi2010b,Sheykhi2007a,Sheykhi2007b,Sheykhi2009,Sheykhi2018,Nojiri2019,Lymperis2018,Sheykhi2019,Saridakis2020,Barrow2021,Saridakis2021,Anagnostopoulos2020,Sheykhi2021,Asghari2021,Sheykhi2022,Sheykhi2022b,Asghari2022,Sheykhi2023,Saleem2023,Sheykhi2023b,Lymperis2021,Drepanou2022,Odintsov2023,Abreu2022,Luciano2022,Coker2023,Okcu2024b,Genarro2022,Odintsov2023b,Nojiri2022,Nojiri2022b,Nojiri2023,Nojiri2023b,Verlinde2011,Cai2010,Shu2010,Sheykhi2010c,Sheykhi2011,Gao2010,Sheykhi2012,Bosso2022,Ling2010,FCai2010,Hendi2011,Basilakos2012,Senay2021,Sefiedgar2017,Feng2018,Abreu2018,Jusufi2023,Padmanabhan2012,Cai2012,Yang2012,Sheykhi2013,Eune2013,Dezaki2015}. For example, Refs. \cite{Padmanabhan2002,Eling2006,Paranjape2006,Kothawala2007,Padmanabhan2007} focus on the derivation of Einstein field equations from the fist law of thermodynamics. Relying on Jacobson's study, Cai and Kim derived the Friedmann equations from the first law of thermodynamics, expressed as  $-dE=TdS$. Here, $-dE$ is interpreted as energy flux that crosses the horizon. However, the temperature is not proportional to the surface gravity of horizon since it is approximately defined for the infinitesimal time at the fixed horizon. Requiring the proportionality between temperature and surface gravity, the first law of horizon is defined by $dE=TdS+WdV$ \cite{Akbar2007}. Here, $E$ and $W$ are understood as  total energy enclosed by volume V and  work density, respectively. Then, many studies targeting the analysis of Friedmann–Lemaitre–Robertson–Walker (FLRW) cosmology from the first law of thermodynamics have been extensively carried out in the literature \cite{Cai2007a,Cai2007b,Cai2008,Sheykhi2010a,Awad2014,Salah2017,Kouwn2018,Okcu2020,Alsabbagh2023,Okcu2024,Luo2023,Das2022BG,Feng2022,Luciano2025,Luciano2021,Luo2025,Sheykhi2010b,Sheykhi2007a,Sheykhi2007b,Sheykhi2009,Sheykhi2018,Nojiri2019,Lymperis2018,Sheykhi2019,Saridakis2020,Barrow2021,Saridakis2021,Anagnostopoulos2020,Sheykhi2021,Asghari2021,Sheykhi2022,Sheykhi2022b,Asghari2022,Sheykhi2023,Saleem2023,Sheykhi2023b,Lymperis2021,Drepanou2022,Odintsov2023,Abreu2022,Luciano2022,Coker2023,Okcu2024b,Genarro2022,Odintsov2023b,Nojiri2022,Nojiri2022b,Nojiri2023,Nojiri2023b}.

Another important advancement in the thermodynamical aspect of gravity can be regarded as Verlinde's entropic gravity \cite{Verlinde2011}. In his approach, gravity is not treated as a fundamental force, but rather as an entropic force that emerges due to the entropy change of the holographic screen. According to Verlinde, the displacement of the test particle from the holographic screen results in the entropic force given by the relation $F\Delta x=T\Delta S$. Here, $\Delta S$ and $T$ are the entropy change and the temperature of the holographic screen, respectively. Considering the holographic screen with Unruh temperature results in Newton's second law. Furthermore,  it is possible to obtain Newton's gravitational law and Einstein field equation from the equipartition law of energy together with the holographic screen. Studies focussing on the derivation of Newton's gravitational law, Einstein field equations, and Friedmann equations in the context of entropic gravity case have been extensively investigated in the literature \cite{Verlinde2011,Cai2010,Shu2010,Sheykhi2010c,Sheykhi2011,Gao2010,Sheykhi2012,Bosso2022,Ling2010,FCai2010,Hendi2011,Basilakos2012,Senay2021,Sefiedgar2017,Feng2018,Abreu2018,Jusufi2023} \footnote{We also call the reader's attention to Padmanabhan's work \cite{Padmanabhan2012}, which regards the spacetime as an emergent structure. Computing the difference between the surface and bulk degrees of freedom, he obtained the Friedmann equations. His notion has also been extensively investigated in the literature \cite{Padmanabhan2012,Cai2012,Yang2012,Sheykhi2013,Eune2013,Dezaki2015}.}. 

Although the thermodynamical aspect of gravity successfully defines the macroscopic description of gravitational systems, it is believed that such a framework may expose significant modifications due to the quantum gravitational effects. In particular, the idea of minimum length scale leads to the modification of the standard Heisenberg Uncertainty Principle (HUP). For example, GUP takes into account  momentum uncertainty correction and may predict the existence of minimal measurable length at Planck scale. On the other hand, the extended uncertainty principle (EUP) takes into account position uncertainty correction and may predict the existence of minimal measurable momentum. A more comprehensive framework occurs as generalised and extended uncertainty principle (GEUP) when both GUP and EUP effects are taken into account. Thus, the modifications of uncertainty principle may play an important role near the Planck scale and large distance scale. Various models of modified uncertainty principle have been investigated in the literature \cite{Maggiore1993,Scardigli1999,Kempf1995,Jizba2010,Bambi2008,Nozari2012,Chung2018,Lake2019,Dabrowski2019,Lake2020,Lake2021,Mureika2019,Du2022,Segreto2023,Ali2009,Ali2011,Vagenas2019,Pedram2012,Nouicer2006}. Modifications of the uncertainty principle may enable a new understanding of black hole thermodynamics \cite{Nouicer2006,Adler2001,Medved2004,Park2007,Nozari2008,Arraut2009,Nozari2012b,Ali2012,Majumder2013,Anacleto2015,Feng2016,Sakalli2016,Kanzi2019,Xiang2009,Scardigli2020a,Hassanabadi2021,Lutfuoglu2021,HassanabadiEUP2019,Bolen2005,Han2008,MoradpourEUP2019,ChungEUP2019,HamilEUP2021,HamilEUP2021b,ChenEUP2019,Sun2018,Ma2018,Zhou2022,Okcu2020b,Okcu2022,Ong2018}. For instance, GUP may prevent black holes from a complete evaporation, resulting in a remnant at the final stage of evaporation \cite{Adler2001}. On the other hand, EUP enables a minimum temperature for the black holes \cite{HassanabadiEUP2019}. Beyond black hole evaporation, the modifications of uncertainty principle also provide rich phase structures and critical phenomena for black holes \cite{Sun2018,Ma2018,Zhou2022,Okcu2020b,Okcu2022}. For example, a reentrant phase transition occurs for the charged AdS black holes in the presence of GUP effects \cite{Sun2018,Ma2018}. In Ref. \cite{Okcu2020b}, we showed that phase transition of van der Waals black hole is physically meaningful for the GUP effects. Moreover, black holes in flat spacetime and a cavity have phase behaviours similar to that of black holes in anti-de Sitter spacetime \cite{Zhou2022,Okcu2022}. 

Beyond the applications of black hole thermodynamics,  modifications of the uncertainty principle have also been investigated within the framework of cosmology \cite{Awad2014,Salah2017,Okcu2020,Alsabbagh2023,Okcu2024,Luo2023,Kouwn2018,Das2022BG,Feng2022,Luciano2025,Luciano2021,Luo2025,Scardigli2011,Nenmeli2021,Das2022,Barca2022,Barca2023,Barca2023b,Bosso2023}. In particular, both the GUP and EUP-modified Friedmann equations have been investigated within the context of gravity-thermodynamics conjecture \cite{Awad2014,Salah2017,Okcu2020,Okcu2024,Luo2023,Alsabbagh2023,Kouwn2018,Das2022BG,Feng2022,Luciano2025,Luciano2021,Luo2025}. For example, GUP models imply a universe without initial singularity \cite{Awad2014,Salah2017,Okcu2020,Alsabbagh2023,Okcu2024}. These models typically predict a  maximum finite energy density  and   minimum apparent horizon at Planck scale, implying a non-singular universe. Furthermore, the thermodynamical consistency of Friedmann equations has been checked by generalised second law (GSL) \cite{Awad2014,Salah2017,Okcu2020,Alsabbagh2023,Okcu2024}. In Ref. \cite{Okcu2024}, we showed the validity of GSL in $\Lambda CDM$ cosmology and found that the redshift parameter is finite and maximum at Planck scale, indicating the absence of initial singularity. Beyond investigations of non-singular nature and thermodynamical consistency, constraining  parameters from observations \cite{Kouwn2018}, gravitational baryogenesis \cite{Das2022BG,Feng2022,Luo2023}, matter perturbation \cite{Luciano2025} and Big Bang Nucleosynthesis \cite{Luciano2021,Luo2025} have also been handled in the framework of modified uncertainty principle \footnote{For a recent review on GUP, please see \cite{Bosso2023}.}.

Modified Friedmann equations are generally derived from modified entropy-area relations within the framework of  gravity-thermodynamics conjecture. In this paper, our aim is to obtain exact Friedmann equations in entropic gravity theory by using the GUP-modified equipartition law of energy.  Modifications of equipartition law of energy may offer a more natural way than modifications of entropy-area relation, as equipartition law of energy is directly linked to the statistical behaviour of the microscopic degrees of freedom of a system. Besides, handling the Friedmann equations from GUP-modified equipartition law of energy may reveal new insight and results that cannot be deduced from the modifications of entropy-area relation . In this context, our goal is also to investigate baryogenesis  for the Firedmann equations from the GUP-modified equipartition law of energy. Baryogenesis is  an important mechanism 
to explain how asymmetry between baryons and antibaryons arises during the early stage of the Universe. It is well known that GUP models have the potential to explain the baryon asymmetry in the radiation-dominated era, because GUP effects can break the thermal equilibrium \cite{Das2022BG,Feng2022,Luo2023}. Consequently, GUP effects
produce a nonzero baryon asymmetry factor $\eta$ characterised by a non-vanishing derivative of Ricci scalar, $\dot{R}$ \cite{Davoudiasl2004}. In our case, the correction terms in equipartition law of energy may be considered as fluctuations. Therefore, these fluctuations due to GUP have potential to break thermal equilibrium in the early stage of the Universe. Accordingly, a nonzero baryon asymmetry factor can be generated. Thus, revisiting the baryon asymmetry is crucial to understanding our approach. In this paper, we will use the simplest form of GUP, which is given by \cite{Maggiore1993, Scardigli1999}
\begin{equation}
\label{GUP}
\Delta x\Delta p\geq\frac{\hbar}{2}\left(1+\frac{\alpha G_{N}}{\hbar c^{3}}\Delta p^{2}\right),
\end{equation}
where $\alpha$ is the dimensionless GUP parameter. This form of GUP is known as quadratic GUP  because the correction term appears as $\Delta p^{2}$. It can be obtained either from gedanken experiments or from suitable deformations of the canonical commutation relations. For example, the gedanken experiments such as Maggiore's analysis \cite{Maggiore1993} of Heisenberg microscope with black holes and Scardigli's analysis \cite{Scardigli1999}, which involves micro-black holes at Planck scale, yield the GUP in Eq. (\ref{GUP}). The same modification is also obtained from modified canonical commutation relation $\left[x,p\right]=i\hbar\left(1+\frac{\alpha G_{N}}{\hbar c^{3}}p^{2}\right)$, which is proposed by Kempf, Mangano and Mann \cite{Kempf1995}.

The paper is organised as follows: In Section \ref{MSEnergy}, we intuitively derive the equipartition law of energy from HUP and thermodynamical properties of the horizon. Then, repeating the same steps for the quadratic GUP, we obtain GUP-modified equipartition law of energy. In Section \ref{ModifiedFriedmann}, employing the modified equipartition law of energy, we obtain the Friedmann equations in the entropic gravity framework. We investigate GUP effects on the Kretschmann scalar. In Section \ref{decelParamSect}, we compute the deceleration parameter and obtain the constraint on the equation of state parameter at the beginning of the Universe. In Section \ref{baryogeSect}, we briefly review gravitational baryogenesis for the standard cosmology. Then, we consider it for GUP case. Moreover, we obtain bounds of GUP parameters from observations. Finally, we discuss our findings in Section \ref{Conlc}.

\section{A heuristic derivation of the equipartition law of energy from the uncertainty principle}

Let us begin to derive the equipartition law from the uncertainty principle and thermodynamical arguments of the horizon. In order to show the procedure, we first consider the HUP. Once we obtain the equipartition law from HUP, the procedure is readily applicable to any modification of uncertainty principle. We assume that the total energy inside the horizon is tantamount to Misner-Sharp energy given by \cite{Akbar2007}
\begin{equation}
\label{MSEnergy}
    E=\frac{c^{4}r_{h}}{2G_{N}},
\end{equation}
where $r_{h}$ is the radius of the horizon. The entropy of the horizon is given by
\begin{equation}
\label{entropy}
S=\frac{k_{B}A}{4\ell_{p}^{2}}=\frac{k_{B}\pi r_{h}^{2}c^{3}}{G_{N}\hbar},
\end{equation}
where $A$ is the area of the horizon screen and we use the Planck length $\ell_{p}=\sqrt{\frac{G_{N}\hbar}{c^{3}}}$. In order to implement the Misner-Sharp energy (\ref{MSEnergy}) and entropy (\ref{entropy}) into HUP, we identify $\Delta x$ with $r_{h}$, i.e., $\Delta x\sim 2r_{h}$ \cite{Adler2001}. Thus, we find the following expressions from Eqs. (\ref{MSEnergy}) and (\ref{entropy})
\begin{equation}
\label{MSEnergy2}
    \Delta x=\frac{4G_{N}E}{c^{4}},
\end{equation}
\begin{equation}
    \label{entropy2}
    \Delta x^{2}=\frac{4G_{N}\hbar S}{k_{B}\pi c^{3}}.
\end{equation}
Furthermore, we identify the momentum uncertainty $\Delta p$ with the horizon temperature $T$. The relation between $\Delta p$ and $T$ is defined by \cite{Adler2001}
\begin{equation}
\label{TempDeltaP}
\Delta p\sim\frac{k_{B}T\gamma}{c},
\end{equation}
where $\gamma$ is proportionality constant.

Now, let us multiply  both sides of HUP $\Delta x\Delta p\geq\frac{\hbar}{2}$ by $\Delta x$. We get
\begin{equation}
\label{HUP2}
\Delta x^{2}\Delta p\geq\frac{\hbar\Delta x}{2}.
\end{equation}
Employing  Eqs. (\ref{MSEnergy2}), (\ref{entropy2}) and (\ref{TempDeltaP}) in the above expression, we obtain $E \lesssim \frac{2ST\gamma}{\pi}$. Here, the energy is bounded from above. This suggests that $E$ cannot exceed $2ST\gamma/\pi$, and at most it can be equal to $2ST\gamma/\pi$. Therefore, we can write 
\begin{equation}
\label{MSEnergy3}
E \approx\frac{2ST\gamma}{\pi}.
\end{equation}
We recall the relation between the number of bits $N$ on the horizon and area, i.e., $N=\frac{c^{3} A}{G_{N}\hbar }$ \cite{Verlinde2011}. Then, the entropy of horizon is given by
\begin{equation}
\label{entropy3}
S=\frac{k_{B}N}{4}.
\end{equation}
Combining the expression in Eq. (\ref{MSEnergy3}) with the above equation, we obtain
\begin{equation}
\label{MSEnergy4}
E\approx\frac{1}{2}\frac{Nk_{B}T\gamma}{\pi}.
\end{equation}
Choosing the proportionality constant as $\gamma=\pi$, we obtain
\begin{equation}
\label{EL}
E=\frac{1}{2}Nk_{B}T.
\end{equation}
This is nothing, but it is just an equipartition law of energy. 

Now, we implement the above procedure into GUP. We first multiply the both sides of GUP (\ref{GUP}) by $\Delta x$. Then, reorganising the GUP (\ref{GUP}) yields the following expression:
\begin{equation}
\label{GUP2}
\Delta x^{2}\Delta p-\frac{\alpha G_{N}\Delta p^{2}\Delta x}{2c^{3}}\geq\frac{\hbar\Delta x}{2}.
\end{equation}
Using  Eqs. (\ref{MSEnergy2}), (\ref{entropy2}) and (\ref{TempDeltaP}) in the above expression, we obtain
\begin{equation}
\label{GUPBoundedE}
E\lesssim\frac{2ST\gamma}{\pi}-\frac{\alpha k_{B}^{2}T^{2}\gamma^{2}}{4}\sqrt{\frac{G_{N}}{\hbar c^{5}}}\sqrt{\frac{4S}{\pi}}.
\end{equation}
This inequality similarly suggests that energy is bounded from above. At most, energy can be equal the right hand side of inequality. Therefore, using Eq. (\ref{entropy3}) in this inequality, we can write
\begin{equation}
\label{GUPEPL}
E\approx\frac{1}{2}\frac{Nk_{B}T\gamma}{\pi}-\frac{\alpha k_{B}^{2}T^{2}\gamma^{2}}{4}\sqrt{\frac{G_{N}}{\hbar c^{5}}}\sqrt{\frac{N}{\pi}}.
\end{equation}
In the limit $\alpha\rightarrow 0$, the above expression must reduce to Eq.(\ref{EL}). Therefore, we find $\gamma=\pi$. Reorganising the above equation and using $\gamma=\pi$, we finally obtain GUP-modified equipartition law of energy \footnote{It is worth noting that the modifications to equipartition law of energy were also been investigated in the entropic gravity. In the limit of low temperature,  the Debye-modified equipartition law was used to obtain Friedmann equations and Einstein field equations \cite{Gao2010,Sheykhi2012}. Furthermore, GUP-modified equipartition law of energy was explored in the framework of entropic gravity. In Ref. \cite{Bosso2022}, Bosso et al. obtained modified equipartition law of energy from GUP-modified density of states. It is given by
\begin{equation}
\label{BossoEtAlEPLE}
E=\frac{1}{2}Nk_{B}T\left[1-\frac{75}{4}\frac{\alpha mk_{B}T}{\left(m_{pl}c\right)^{2}}\right], \nonumber
\end{equation}
where m and $m_{pl}$ are mass of test particle and Planck mass, respectively. This approach results in system-dependent expression of equipartition law of energy. In contrast, our approach is more heuristic, and based on the thermodynamical arguments of horizon with model-independent framework. Thus, the correction term in Eq. (\ref{GUPEPL2}) only depends on  thermodynamical properties of horizon, such as $T$ and $N$.}
\begin{equation}
\label{GUPEPL2}
E=\frac{1}{2}Nk_{B}T\left(1-\frac{\alpha\sqrt{\pi^{3}}}{2}\sqrt{\frac{G_{N}}{\hbar c^{5}}}\frac{k_{B}T}{\sqrt{N}}\right).
\end{equation}

In order to understand the effects of correction term, we consider the classical limit (low energy limit) and quantum gravity limit (high energy limit) \cite{Bolen2005}. In the classical limit, the second term must be negligible. The classical limit is given by
\begin{equation}
\label{classicalLimit}
\frac{\alpha\sqrt{\pi^{3}}}{2\sqrt{N}}\frac{k_{B}T}{E_{P}}\ll1 \qquad\Longrightarrow\qquad \frac{k_{B}T}{E_{P}}\ll\frac{2\sqrt{N}}{\alpha\sqrt{\pi^{3}}},
\end{equation}
where $E_{P}$ is the Planck energy, and we use $E_{P}=\sqrt{\frac{\hbar c^{5}}{G_{N}}}$. The classical limit  is valid when the thermal energy of horizon $k_{B}T$ is much smaller than the Planck energy, i.e., $k_{B}T\ll E_{P}$, and the number of bits is very large, i.e., $N\rightarrow\infty$. This latter condition resembles the thermodynamic limit $\tilde{N}\rightarrow\infty$ where $\tilde{N}$ is the number of particles in the conventional systems. Thus, we may refer to this limit as thermodynamical limit of the number of bits. The quantum gravity limit is given by
\begin{equation}
\label{QGLimit}
\frac{\alpha\sqrt{\pi^{3}}}{2\sqrt{N}}\frac{k_{B}T}{E_{P}}\sim1 \qquad\Longrightarrow\qquad \frac{k_{B}T}{E_{P}}\sim\frac{2\sqrt{N}}{\alpha\sqrt{\pi^{3}}}.
\end{equation}
This limit is valid when the thermal energy of horizon is comparable to the Planck energy, i.e., $k_{B}T\approx E_{P}$, and the number of bits is very small. This latter condition resembles the case of small number of particles in the conventional systems \cite{Greiner}. It is well known that the fluctuations can be quite large in a small spatial region of the conventional systems. Similarly, the quantum fluctuations can also be considered sufficiently large at the Planck scale.

\section{Friedmann equations from the entropic gravity}
\label{ModifiedFriedmann}

Let us commence a brisk review of the fundamental elements of the FLRW universe. The line element is expressed by \cite{Cai2005}
\begin{equation}
 \label{lineElement}
 ds^{2}=h_{ab}dx^{a}dx^{b}+\tilde{r}^{2}d\Omega^{2},
 \end{equation}
where $\tilde{r}=a(t)r$, the coordinates $x^0=ct$, $x^1=r$ and two-dimensional  metric $ h_{ab}= diag\left(-1,a^{2}/(1-kr^{2})\right)$. Here $a(t)$ is the scale factor and  the spatial curvature constant $k=$ $-1$, $0$ and $1$ correspond to open, flat and closed universe, respectively. The line element of unit two-sphere is given by $d\Omega^{2}=d\theta^{2}+\sin^{2} \theta \,d\varphi^{2} .$ The apparent horizon is obtained from the relation \cite{Cai2005}
\begin{equation}
\label{DefiniOfRa}
h^{\mu\nu}\partial_{\mu}\tilde{r_{A}}\partial_{\nu}\tilde{r_{A}}=0.
\end{equation}
The apparent horizon is given by
 \begin{equation}
 \label{apparentHor}
 \tilde{r_{A}}=ar=\frac{c}{\sqrt{H^{2}+kc^{2}/a^{2}}},
 \end{equation}
where $H(t)=\dot{a}/a$ is the Hubble parameter, and dot represents the derivative with respect to time. We define the energy and matter of the Universe as a perfect fluid. The corresponding energy-momentum tensor is defined by
 \begin{equation}
 \label{energyMomentumTensor}
 \mathcal{T}_{\mu\nu} = (\rho+p)\frac{u_{\mu}u_{\nu}}{c^{2}} + p g_{\mu\nu},
 \end{equation}
 where $\rho$, $p$ are energy density and pressure of the fluid, $g_{\mu\nu}$ is the metric tensor,  and $u^{\mu}$ is four-velocity of the fluid. In the matter-dominated case, one may write $\rho = \rho_{m} c^{2}$, where $\rho_{m}$ denotes the mass density. The continuity equation is obtained from the conservation of energy-momentum tensor $\nabla_{\mu}\mathcal{T}^{\mu\nu}=0$ given by \footnote{In this work, the GUP modification arises from horizon thermodynamics where the entropy and  temperature are associated with the geometrical properties of the horizon such as area and acceleration, respectively. Therefore, GUP modification is only induced due to the geometrical properties of the horizon, without touching the energy-momentum tensor. Consequently, we consider the energy-momentum tensor to be preserved as in the standard case.}
  \begin{equation}
 \label{continuityEqu}
 \dot{\rho}+3H(\rho+p)=0.
 \end{equation}
Following the methods in Refs.\cite{Verlinde2011,Cai2010}, we are going to derive the GUP-modified Friedmann equations from entropic gravity. We begin to consider a holographic screen on the boundary $\mathcal{\partial V}$ of a compact spatial region $\mathcal{V}$. The number of bits on the holographic screen is defined by Eq. (\ref{entropy3}). We suppose that the total energy of the screen follows the equipartition law of energy (\ref{EL}). In our case, we employ the GUP-modified equipartition law of energy (\ref{GUPEPL2}). The  temperature of holographic screen corresponds to Unruh temperature given by \cite{Verlinde2011}
\begin{equation}
\label{unruhTemperature}
T=-\frac{\hbar a_{r}}{2\pi k_{B}c}=-\frac{\hbar\ddot{a}r}{2\pi k_{B}c},
\end{equation}
where $a_{r}$ is the acceleration. It is given by \cite{Cai2010}
\begin{equation}
\label{accl}
a_{r}=-\frac{d^{2}\tilde{r_{A}}}{dt^{2}}=-\ddot{a}r.
\end{equation}
Combining Eqs. (\ref{entropy}), (\ref{entropy3}), (\ref{apparentHor}) and (\ref{unruhTemperature}) with (\ref{GUPEPL2}), we obtain the following expression
\begin{equation}
\label{eqlFr}
E=-\frac{\ddot{a}a^{2}r^{3}c^{2}}{G_{N}}\left(1+\frac{\alpha}{8}\frac{G_{N}\hbar}{c^{5}}\frac{\ddot{a}}{a}\right).
\end{equation}
To get the modified Friedmann equations, we also consider the active gravitational mass $\mathcal{M}$ inside the horizon. It is defined by \cite{Cai2010} 
\begin{equation}
\label{KomarMass}
\mathcal{M}=2\int_{\mathcal{V}}dV\left(\mathcal{T}_{\mu\nu}-\frac{1}{2}\mathcal{T}g_{\mu\nu}\right)u^{\mu}u^{\nu}=\frac{4\pi\widetilde{r_{A}}^{3}}{3c^{2}}(\rho+3p).
\end{equation}
Using the assumption $\mathcal{M}c^2=E$ with Eqs. (\ref{eqlFr}) and (\ref{KomarMass}), we find the acceleration equation
\begin{equation}
\label{QuadraticFriedmann}
\frac{\ddot{a}}{a}\left(1+\frac{\alpha G_{N}\hbar}{8c^{5}}\frac{\ddot{a}}{a}\right)=-\frac{4\pi G_{N}}{3c^{2}}(\rho+3p).
\end{equation}
Interestingly, the quantum gravity correction appears as a quadratic term of $\ddot{a}/a$, that is, we may infer that  quantum gravity directly affects the acceleration of the Universe. The above equation yields two solutions.
\begin{equation}
\label{DynamicalEqRoots}
\frac{\ddot{a}}{a}=-\frac{4c^{5}}{G_{N}\hbar\alpha}\left(1\pm\sqrt{\frac{3c^{7}-2G_{N}^{2}\hbar\pi\alpha(\rho+3p)}{3c^{7}}}\right).
\end{equation}
We choose the  solution with negative sign in front of the square root since it reduces to standard acceleration equation in the limit $\alpha\rightarrow0$. In order to get a real solution of $\ddot{a}$, $\rho$ must have a maximum value given by
\begin{equation}
\label{rhoMax}
\rho_{max}=\frac{3}{2\pi\alpha(1+3\omega)}\frac{c^{7}}{G_{N}^{2}\hbar}=\frac{3\rho_{p\ell}c^{2}}{2\pi\alpha(1+3\omega)}\approx\frac{2.214\times10^{113}kgm^{-1}s^{-2}}{\alpha(1+3\omega)},
\end{equation}
where we use Planck density $\rho_{p\ell}=c^{5}/(G_{N}^{2}\hbar)$ and equation of state $p=\omega \rho$. This result implies that the Universe started to expand at a maximum and finite energy density. In other words, the singularity is removed at the beginning of the Universe. In fact, a universe without initial singularity is the generic feature of GUP models \cite{Awad2014,Salah2017,Okcu2020,Alsabbagh2023,Okcu2024}. The positivity of maximum energy density depends on the GUP parameter $\alpha$ and the equation of state parameter $\omega$. Generally, the minimum apparent horizon and maximum energy density are defined when the GUP parameter is positive \cite{Awad2014,Salah2017,Okcu2020,Alsabbagh2023,Okcu2024}. As can be seen from the above equation, the condition $1+3\omega>0$ implies  $\alpha>0$ for the positivity of energy density. Interestingly, the condition $1+3\omega<0$ allows a negative GUP parameter. This peculiar behaviour arises due to dependence of maximum energy density on both $\omega$ and $\alpha$. Namely, both parameters play a decisive role on the positivity of energy density. However, $\rho_{max}$ diverges at $\omega=-1/3$. At this point, the GUP-induced maximum energy density is not defined. In fact, $\omega=-1/3$ separates two regions in which the sign of the GUP parameter switches. Thus, we may infer that $\omega=-1/3$ acts as critical point. The first hint for the sign of $\alpha$ can be seen from Eq. (\ref{DynamicalEqRoots}). At the maximum energy density, the dynamical equation is given by $\frac{\ddot{a}}{a}=-\frac{4c^{5}}{G_{N}\hbar \alpha}$. It is clear that $\alpha$ must be negative  because the initial acceleration requires the condition $\ddot{a}/a>0$.

Let us center our interest on the derivation of the first Friedmann equation. Multiplying $2\dot{a}a$ on both sides of Eq. (\ref{DynamicalEqRoots}), the integral form of Eq. (\ref{DynamicalEqRoots}) is given by
\begin{equation}
\label{firstFriedmannInt1}
\int\frac{d\dot{a}^{2}}{dt}dt=-\frac{8c^{5}}{\alpha G_{N}\hbar}\int a\frac{da}{dt}dt+\frac{8c^{5}}{\alpha G_{N}\hbar}\int a\sqrt{\frac{3c^{7}-2G_{N}^{2}\hbar\pi\alpha(\rho+3p)}{3c^{7}}}\frac{da}{dt}dt.
\end{equation}
We choose again the equation of state as $p=\omega \rho$. To solve the above integral, we must know the relation between the energy density $\rho$ and the scale factor $a$. Employing $p=\omega \rho$ in Eq. (\ref{continuityEqu}) yields
\begin{equation}
\label{solutionContiEq}
\rho=\rho_{0}a^{-3(1+\omega)},
\end{equation}
where $\rho_{0}$ is the integration constant, and it can be identified with the energy density of the fluid at the present epoch when $a_{0}=1$. Using the above expression in Eq. (\ref{firstFriedmannInt1}), the integral is given by
\begin{equation}
\label{firstFriedmannInt2}
\int\frac{d\dot{a}^{2}}{dt}dt=-\frac{8c^{5}}{\alpha G_{N}\hbar}\int a\frac{da}{dt}dt+\frac{8c^{5}}{\alpha G_{N}\hbar}\int a\sqrt{\frac{3c^{7}-2G_{N}^{2}\hbar\pi\alpha(1+3\omega)\rho_{0}a^{-3(\omega+1)}}{3c^{7}}}\frac{da}{dt}dt.
\end{equation}
Solving this integral yields the first Friedmann equation as follows:
\begin{equation}
\label{firstFriedmannEq}
\frac{\dot{a}^{2}}{a^{2}}+\frac{kc^{2}}{a^{2}}=-\frac{4c^{5}}{\alpha G_{N}\hbar}\left(1-_{2}F_{1}\left[\frac{-1}{2},\frac{-2}{3(1+\omega)},\frac{1+3\omega}{3(1+\omega)};\frac{2\pi\alpha(1+3\omega)\rho}{3}\frac{G_{N}^{2}\hbar}{c^{7}}\right]\right)
\end{equation}
where $_{2}F_{1}\left[k,l,m;n\right]$ is the hypergeometric function. In order to understand the behaviour of the first Friedmann equation at $\rho=\rho_{max}$, we employ a convenient transformation to the hypergeometric function in Eq. (\ref{firstFriedmannEq}). This transformation is given as follows \cite{Middleton2011,Abramowitz1965}:
\begin{align}
{}_2F_{1}[u,l,m;n] &= 
\frac{\Gamma(m)\,\Gamma(u+l-m)}{\Gamma(u)\,\Gamma(l)}\,
(1-n)^{\,m-k-l}\,
{}_2F_{1}\!\left[m-u,\,m-l,\,m-u-l+1;\,1-n\right] \nonumber \\
&\quad+\;
\frac{\Gamma(m)\,\Gamma(m-u-l)}{\Gamma(m-u)\,\Gamma(m-l)}\,
{}_2F_{1}\!\left[u,\,l,\,u+l-m+1;\,1-n\right],
\end{align}
where $\Gamma$ denotes the gamma function. Applying this transformation to hypergeometric function in Eq. (\ref{firstFriedmannEq}), we find
\begin{multline}
{}_{2}F_{1}\!\left[-\tfrac{1}{2},-\tfrac{2}{3(w+1)},\tfrac{3w+1}{3(w+1)};\tfrac{\rho}{\rho_{\max}}\right]
= \frac{\Gamma\!\left(\tfrac{3w+1}{3(w+1)}\right)\Gamma\!\left(-\tfrac{3}{2}\right)}
{\Gamma\!\left(-\tfrac{1}{2}\right)\Gamma\!\left(-\tfrac{2}{3(w+1)}\right)}
\left(1-\tfrac{\rho}{\rho_{\max}}\right)^{\tfrac{3}{2}}\\
\times {}_{2}F_{1}\!\left[\tfrac{9w+5}{6(w+1)},1,\tfrac{5}{2};1-\tfrac{\rho}{\rho_{\max}}\right]
+\frac{\Gamma\!\left(\tfrac{3w+1}{3(w+1)}\right)\Gamma\!\left(\tfrac{3}{2}\right)}
{\Gamma\!\left(\tfrac{9w+5}{6(w+1)}\right)}\,
{}_{2}F_{1}\!\left[-\tfrac{1}{2},-\tfrac{2}{3(w+1)},-\tfrac{1}{2};1-\tfrac{\rho}{\rho_{\max}}\right].
\end{multline}
Moreover, this function can be simplified at $\rho=\rho_{max}$. The prefactor in the first term vanishes at $\rho=\rho_{max}$. Thus the first term does not give any contribution. The hypergeometric function in the second term reduces to unity at $\rho=\rho_{max}$ since ${}_{2}F_{1}\!\left[k,l,m,0\right]=1$ \cite{Abramowitz1965}. Therefore, the hypergeometric function in Eq. (\ref{firstFriedmannEq}) is simply expressed in terms of gamma functions as
\begin{equation}
   {}_{2}F_{1}\!\left[-\tfrac{1}{2}, -\tfrac{2}{3(w+1)}, \tfrac{3w+1}{3(w+1)}; 1\right]
= \frac{\sqrt{\pi}}{2}\,
\frac{\Gamma\!\left(\tfrac{3w+1}{3(w+1)}\right)}
{\Gamma\!\left(\tfrac{9w+5}{6(w+1)}\right)}. 
\end{equation}
Substituting the above expression into Eq. (\ref{firstFriedmannEq}) and using $H=\dot{a}/a$ , we find the maximum Hubble parameter $H_{max}$ at $\rho=\rho_{max}$. At this point, we focus on the flat case $k=0$ for simplicity. $H_{max}$ is given by
\begin{equation}
H_{max}=\mp\sqrt{\frac{-4c^{5}}{G_{N}\hbar\alpha}\left(1-\frac{\sqrt{\pi}}{2}\frac{\Gamma\left(\frac{3\omega+1}{3\left(\omega+1\right)}\right)}{\Gamma\left(\frac{9\omega+5}{6\left(\omega+1\right)}\right)}\right)}.
\end{equation}
Here, we consider the positive branch of solution since it corresponds to an expanding universe. At the maximum energy density, the Hubble parameter is finite. This result clearly signals the avoidance of the initial singularity. For $\omega=-1$, $H_{max}$ becomes ill-defined. However, in the next section, we will show the pathological behaviour of $\omega\leq-1$ in the analysis of the deceleration parameter.  In order to see the effects of GUP on $H_{max}$, we show  $H_{max}$ for different values of $\alpha$ in Tables \ref{tablo1} and \ref{tablo2}. The chosen values of $\omega$ illustrate the behaviours of positive and negative GUP parameters. As can be seen from the tables, the values of $H_{max}$ depend on the magnitude of the GUP parameter $\left|\alpha\right|$. $H_{max}$ decreases for the increasing value of $\left|\alpha\right|$.
\begin{table}[ht!]
\centering
\begin{tabular}{c|cccccc}
\hline
$\alpha$ & $0.2$ & $0.4$ & $0.6$ & $0.8$ & $1.0$ & $1.2$ \\
\hline
$H_{max}$ & $3.38$ & $2.39$ & $1.95$ & $1.69$ & $1.51$ & $1.38$ \\
\hline
\end{tabular}
\caption{Values of $H_{max}$ for $w=\tfrac{1}{3}$ and different positive $\alpha$. We use $G_{N}=\hbar=c=1$.}
\label{tablo1}
\end{table}
\begin{table}[ht!]
\centering
\begin{tabular}{c|cccccc}
\hline
$\alpha$ & $-1.2$ & $-1.0$ & $-0.8$ & $-0.6$ & $-0.4$ & $-0.2$ \\
\hline
$H_{max}$ & $1.26$ & $1.38$ & $1.54$ & $1.78$ & $2.18$ & $3.08$ \\
\hline
\end{tabular}
\caption{Values of $H_{max}$ for $w=-0.6$ and different $\alpha$. We use $G_{N}=\hbar=c=1$.}
\label{tablo2}
\end{table}

\begin{figure}
\centering
\includegraphics[width=8.18cm]{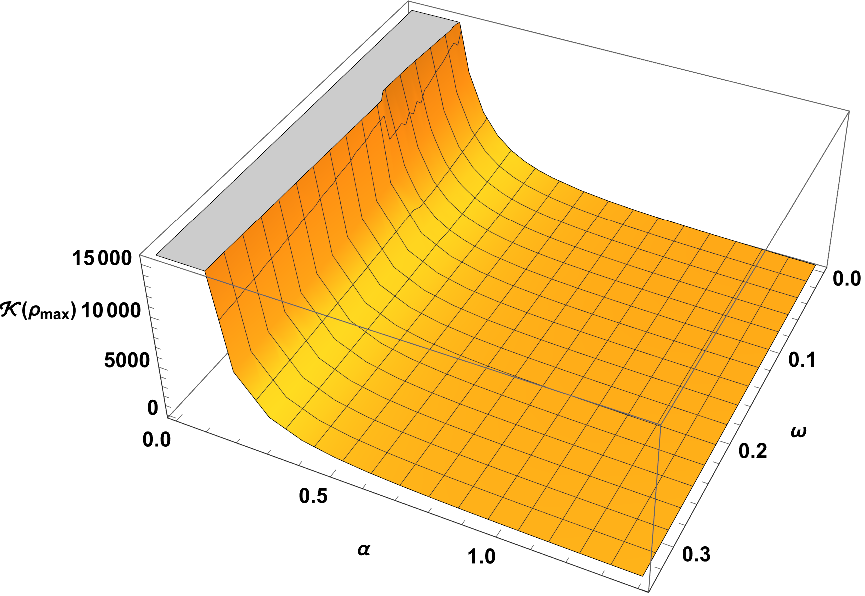}
\includegraphics[width=8.18cm]{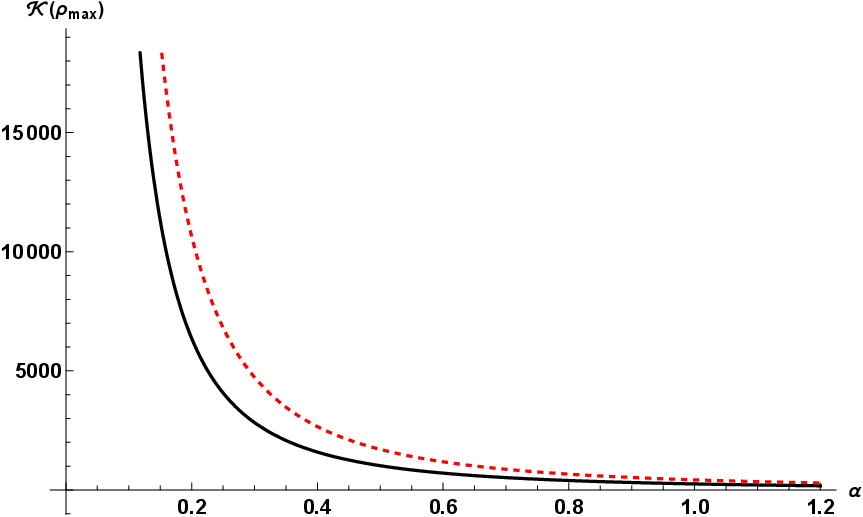}
\caption{(a) Kretschmann curvature scalar at maximum density vs GUP parameter and equation of state parameter. (b) Kretschmann curvature scalar at maximum density vs GUP parameter. Black solid and red dashed lines corresponds to $\omega=1/3$ and $\omega=0$, respectively. We use $G_{N}=\hbar=c=1$.}
\label{KrestchFig}%
\end{figure}

\begin{figure}
\centering
\includegraphics[width=8.18cm]{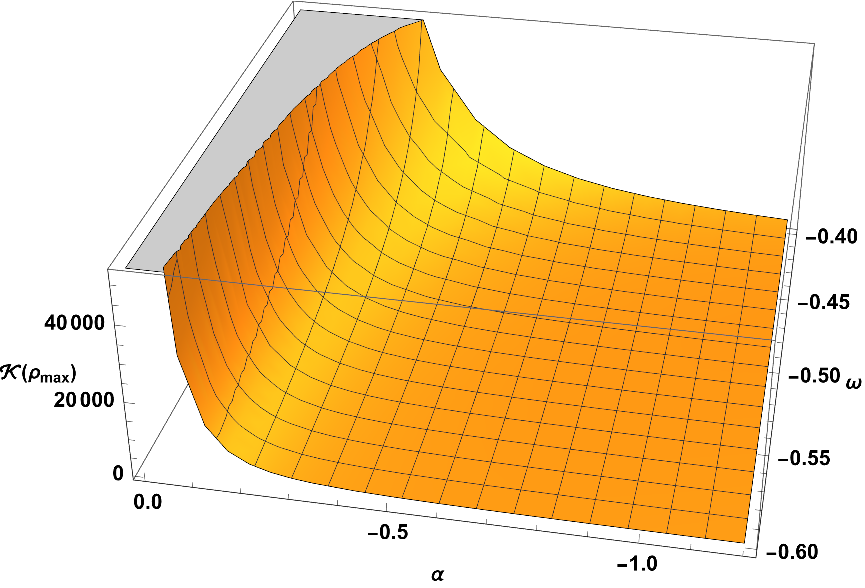}
\includegraphics[width=8.18cm]{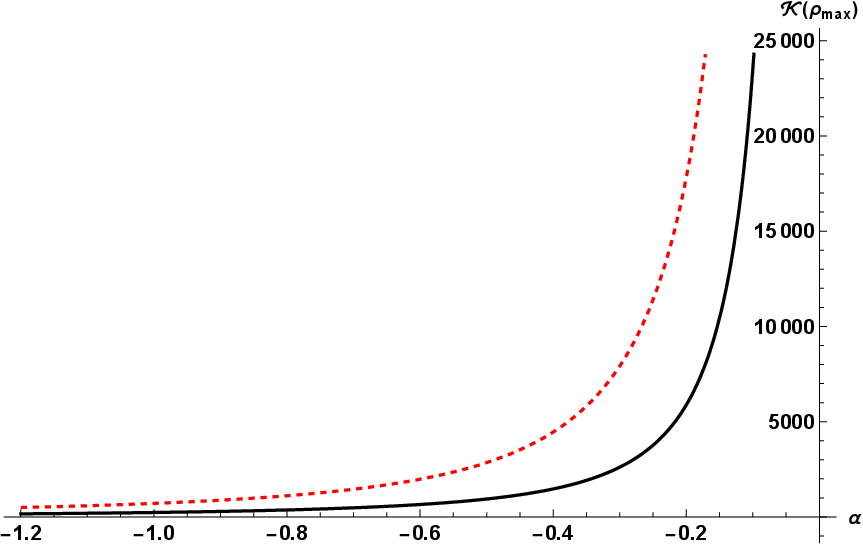}
\caption{(a) Kretschmann curvature scalar at maximum density vs GUP parameter and equation of state parameter. (b) Kretschmann curvature scalar at maximum density vs GUP parameter. Black solid and red dashed lines corresponds to $\omega=-0.6$ and $\omega=-0.5$, respectively. We use $G_{N}=\hbar=c=1$.}
\label{KrestchFig2}%
\end{figure}

Another useful way to show that the Universe has no initial singularity is to investigate the Kretschmann curvature scalar. If the Kretschmann curvature scalar is finite at the beginning of the Universe, this implies the absence of the initial singularity. For FLRW metric, it is given by
\begin{equation}
\label{KretsCurvScal}
\mathcal{K}=R^{\mu\nu\rho\gamma}R_{\mu\nu\rho\gamma}=12\frac{k^{2}c^{4}+2kc^{2}\dot{a}^{2}+\dot{a}^{4}+\ddot{a}^{2}a^{2}}{a^{4}c^{4}}=\frac{12}{c^{4}}\left[\left(\frac{\dot{a}^{2}}{a^{2}}+\frac{kc^{2}}{a^{2}}\right)^{2}+\frac{\ddot{a}^{2}}{a^{2}}\right],
\end{equation}
where $R_{\mu\nu\rho\gamma}$ is the Riemann tensor. Employing the first and second Friedmann equations in Eq. (\ref{KretsCurvScal}), we find the Kretschmann curvature scalar at the maximum energy density as follows:
\begin{equation}
\label{KretsCurvScal2}
\mathcal{K}\left(\rho_{max}\right)=\frac{192c^{6}}{G_{N}^{2}\hbar^{2}\alpha^{2}}\left(\left(1-_{2}F_{1}\left[-\frac{1}{2},-\frac{2}{3(1+\omega)},\frac{1+3\omega}{3(1+\omega)};1\right]\right)^{2}+1\right)
\end{equation}
In Figs. (\ref{KrestchFig}) and (\ref{KrestchFig2}), we present the Kretschmann curvature scalar with respect to $\alpha$ and $\omega$. As can be seen in both figures, $\mathcal{K}(\rho_{max})$ is finite, that is, implying a regular spacetime at the beginning of Universe. One can see that the intensity of $\mathcal{K}\left(\rho_{max}\right)$ decreases when $|\alpha|$ increases. For small values of $|\alpha|$, the increase in $\mathcal{K}\left(\rho_{max}\right)$ indicates the strong curvature of spacetime at maximum density. When $|\alpha|$ increases, $\mathcal{K}\left(\rho_{max}\right)$ dramatically decreases. These results indicate how GUP effects play a vital role at the beginning of the Universe, namely, effectively removing the initial singularity. The curvature behaviour of spacetime in the early Universe depends on the magnitude of GUP parameter $|\alpha|$.   For the limit $\alpha \rightarrow0$, the Kretschmann curvature scalar diverges. 


Using the equation of state $p=\omega \rho$ again, one can write the first Friedmann equation in terms of $\rho$ and $p$
\begin{equation}
\label{firstFriedmannEq2}
\frac{\dot{a}^{2}}{a^{2}}+\frac{kc^{2}}{a^{2}}=-\frac{4c^{5}}{G_{N}\hbar\alpha}\left(1-_{2}F_{1}\left[\frac{-1}{2},\frac{-2\rho}{3(\rho+p)},\frac{\rho+3p}{3(\rho+p)};\frac{2\pi\alpha(\rho+3p)}{3}\frac{G_{N}^{2}\hbar}{c^{7}}\right]\right).
\end{equation}
When GUP effects are small, the Friedmann equations can be expanded as
 \begin{eqnarray}
 \label{FriedmanEqExpanded}
 \frac{\dot{a}^{2}}{a^{2}}+\frac{kc^{2}}{a^{2}}=\frac{8\pi G_{N}\rho}{3c^{2}}+\frac{2\alpha\pi^{2}G_{N}^{3}\hbar}{9c^{9}}\frac{\rho(\rho+3p)^{2}}{2\rho+3p}+\mathcal{O}\left(\alpha^{2}\right),\nonumber\\
 \frac{\ddot{a}}{a}=-\frac{4\pi G_{N}}{3c^{2}}\left(\rho+3p\right)-\frac{2\alpha\pi^{2}G_{N}^{3}\hbar}{9c^{9}}\left(\rho+3p\right)^{2}+\mathcal{O}\left(\alpha^{2}\right),
 \end{eqnarray}
where the second terms in both equations correspond to the first order quantum gravity correction due to the GUP. Interestingly, this term shares some similarities with braneworld cosmology in which quadratic corrections in $\rho$ appear \cite{Binetruy2000,Binetruy2000b,Ida2000,Shiromizu2000,Ichiki2002,Bernal2020}. To make this case more transparent, we can write the first Friedmann equation (\ref{FriedmanEqExpanded}) in the following form:
\begin{equation}
\label{braneAnalogy}
\frac{\dot{a}^{2}}{a^{2}}=\frac{8\pi G_{N}\rho}{3c^{2}}\left(1+\frac{\left(1+3\omega\right)\rho}{8\left(2+3\omega\right)\rho_{max}}\right),
\end{equation}
where we use $\rho_{max}$ (\ref{rhoMax}), $p=\omega\rho$, and take $k=0$ for simplicity. Moreover, defining $\chi=\frac{8(2+3\omega)\rho_{max}}{(1+3\omega)}$, the above equation is written by
\begin{equation}
\label{braneAnalogy2}
\frac{\dot{a}^{2}}{a^{2}}=\frac{8\pi G_{N}\rho}{3c^{2}}\left(1+\frac{\rho}{\chi}\right).
\end{equation}
This makes the analogy with braneworld cosmology explicit. In the braneworld cosmology, the first Friedmann equation for a spatially flat case is given by  $\frac{\dot{a}^{2}}{a^{2}}=\frac{8\pi G_{N}\rho}{3c^{2}}\left(1+\frac{\rho}{\sigma}\right)$, where $\sigma$ is the brane tension \cite{Bernal2020}. Comparing this with the above equation, one can see that $\chi$ plays the role of $\sigma$. Therefore, we may infer this correspondence that the GUP-induced quadratic correction in $\rho$ mimics the $\rho^{2}$ term in braneworld cosmology. Interestingly, this correction stems from quantum-gravity effects due to the GUP, without invoking higher-dimensional geometry. It is worth noting that this resemblance holds only for the quadratic term in energy density. Other contributions in braneworld cosmology such as cosmological constant and dark radiation terms do not have any counterparts in our case \cite{Ichiki2002}.

\section{Deceleration parameter}
\label{decelParamSect}

We now turn our attention to the deceleration parameter to understand GUP effects on the early Universe. The deceleration parameter is crucial for determining whether a universe is an accelerated or a decelerated phase. It is defined by
\begin{equation}
\label{decelParam}
q=\frac{-a\ddot{a}}{\dot{a}^{2}},
\end{equation}
where the case $q>0$ implies the decelerated phase, while the case $q<0$ implies the accelerated phase. Choosing $p=\omega \rho$ and employing the GUP-modified Friedmann equations (\ref{DynamicalEqRoots}) and (\ref{firstFriedmannEq})  in Eq. (\ref{decelParam}), we find
\begin{equation}
\label{decelParam1}
q=\frac{1-\sqrt{\frac{3c^{7}-2G_{N}^{2}\hbar\pi\alpha(3\omega+1)\rho}{3c^{7}}}}{_{2}F_{1}\left[\frac{-1}{2},\frac{-2}{3(1+\omega)},\frac{1+3\omega}{3(1+\omega)};\frac{2\pi\alpha(1+3\omega)\rho}{3}\frac{G_{N}^{2}\hbar}{c^{7}}\right]-1}
\end{equation}
for the flat case $k=0$. For the tiny GUP effects, the deceleration parameter can be expanded as
\begin{equation}
\label{decelParaExpan}
q=\frac{3\omega+1}{2}+\frac{\hbar G_{N}^{2}\pi\left(\omega+1\right)\left(3\omega+1\right)^{2}\alpha\rho}{8c^{7}\left(3\omega+2\right)}+\mathcal{O}\left(\alpha^{2}\right)
\end{equation}
where the first and second terms correspond to standard case and the first order quantum correction, respectively. Using Eq. (\ref{solutionContiEq}) within the deceleration parameter (\ref{decelParam1}), we get
\begin{equation}
\label{decelParam2}
\begin{aligned}
  q &=\frac{1-\sqrt{\frac{3c^{7}-2G_{N}^{2}\hbar\pi\alpha(3\omega+1)\rho_{0}a^{-3(\omega+1)}}{3c^{7}}}}{_{2}F_{1}\left[\frac{-1}{2},\frac{-2}{3(1+\omega)},\frac{1+3\omega}{3(1+\omega)};\frac{2\pi\alpha(1+3\omega)\rho_{0}a^{-3(\omega+1)}}{3}\frac{G_{N}^{2}\hbar}{c^{7}}\right]-1}\\
&=\frac{1-\sqrt{\frac{3c^{7}-2G_{N}^{2}\hbar\pi\alpha(3\omega+1)\rho_{0}(1+z)^{3(\omega+1)}}{3c^{7}}}}{_{2}F_{1}\left[\frac{-1}{2},\frac{-2}{3(1+\omega)},\frac{1+3\omega}{3(1+\omega)};\frac{2\pi\alpha(1+3\omega)\rho_{0}(1+z)^{3(\omega+1)}}{3}\frac{G_{N}^{2}\hbar}{c^{7}}\right]-1}.  
\end{aligned}
\end{equation}
In the second line, $z$ is the redshift parameter and we use the formula $z+1=\frac{a_{0}}{a}$ by choosing the present scale factor $a_{0}=1$. The deceleration parameter yields the minimum scale factor $a_{min}$ and the maximum redshift parameter $z_{max}$ as
\begin{equation}
\label{amin}
a_{min}=\left(\frac{2\pi\alpha(3\omega+1)\rho_{0}}{3\rho_{p\ell}c^{2}}\right)^{\frac{1}{3(\omega+1)}}=\left(\frac{\rho_{0}}{\rho_{max}}\right)^{\frac{1}{3(\omega+1)}},
\end{equation}
\begin{equation}
\label{zmax}
z_{max}=\left(\frac{3\rho_{p\ell}c^{2}}{2\pi\alpha(3\omega+1)\rho_{0}}\right)^{\frac{1}{3(\omega+1)}}-1=\left(\frac{\rho_{max}}{\rho_{0}}\right)^{\frac{1}{3(\omega+1)}}-1,
\end{equation}
respectively. Here we use the definition of $\rho_{max}$ (\ref{rhoMax}) in the second expressions of Eqs. (\ref{amin}) and (\ref{zmax}). The existence of minimum scale factor and maximum redshift parameter clearly indicates a universe without an initial singularity. The positivity of both parameters depends on $\alpha$ and $\omega$. The conditions $\omega>-1/3$ and $\alpha>0$ ensure the positivity of $a_{min}$ and $z_{max}$. However, for $\alpha<0$, one must be careful. Although the condition $\omega<-1/3$ is sufficient to guarantee the positivity of $a_{min}$,  one must consider $\omega$ in the range $(-1,-1/3)$. Otherwise, if $w$ were in the range $\omega\leq-1$, $a_{min}$ would become greater than $a_{0}=1$ due to the negative exponent. Similarly, $z_{max}$ would approach $-1$ since the contribution from the first term would be tiny due to the negative exponent. These results correspond to future eras of the Universe rather than its past. Thus, they are physically meaningless. For the negative values of $\alpha$, $\omega$ must be in the range $-1<\omega<-1/3$. Beyond the standard cosmology, we can obtain some constraints on $\omega$ from the deceleration parameter at the beginning of the Universe. These constraints emerge due to the GUP effects. 

\section{Baryogenesis}
\label{baryogeSect}

One of the mysterious problems in modern cosmology is baryon asymmetry, which is the prevalence of matter over antimatter in the Universe. Although the number of matter and antimatter in the Universe was initially considered to be equal, various observations \cite{PDG,Planck,Riotto1998,WMAP,Cyburt2004} have refuted this thought and  indicated an excess of matter over antimatter. For the baryon asymmetry to occur, three conditions must be satisfied: 1) Violation of baryon number, 2) Violation of C and CP, 3) Breaking of thermal equilibrium. These conditions are called Sakharov conditions \cite{Sakharov}. Since then, numerous theories have been suggested to account for the baryon asymmetry of the Universe \cite{Canetti2012,Lambiase2013,Cui2015,Oikonomou2016,Oikonomou2016b,Nozari2018,Bhattacharjee2020}. Especially, considering the dynamical breaking of CPT, Davoudiasl et al. \cite{Davoudiasl2004} suggested gravitational baryogenesis mechanism, which indicates viable baryon asymmetry in thermal equilibrium. The interaction defined by the coupling between the derivative of Ricci scalar and the baryon current $J^{\mu}$ satisfies the first two Sakharov conditions, while the last condition is not satisfied. Moreover, the gravitational baryogenesis cannot generate baryon asymmetry for the radiation-dominated era of the Universe in the framework of standard cosmology. In order to solve this issue, the gravitational baryogenesis mechanism has been investigated within the context of modified gravity theories \cite{Oikonomou2016,Oikonomou2016b,Nozari2018,Bhattacharjee2020}. As mentioned in the Introduction, the GUP effects can break the thermal equilibrium, thus satisfying the third condition \cite{Das2022BG,Feng2022,Luo2023}. Combining GUP effects with gravitational baryogenesis provides a satisfactory explanation for baryon asymmetry in the radiation-dominated era. In Ref. \cite{Das2022BG},  Das et al.  already investigated the baryon asymmetry for the GUP-modified Friedmann equations, which were obtained from the first law of thermodynamics at the apparent horizon using the GUP-modified entropy area relation. In our case, we revisit the baryon asymmetry to investigate whether the GUP modification to the equipartition law of energy can break thermal equilibrium, thus satisfying the third Sakharov condition.

The interaction, which breaks the CPT dynamically in an expanding universe, is defined by coupling between the baryon current $J^{\mu}$ and the derivative of Ricci scalar $R$ \cite{Davoudiasl2004}
\begin{equation}
\label{BA1}
\frac{1}{M_{*}^{2}}\int dx^{4}\sqrt{-g}\left(\partial_{\mu}R\right)J^{\mu},
\end{equation}
where $M_{*}$ and $g$ are the cutt-off scale of the effective theory and determinant of metric tensor, respectively \footnote{In this section, we use the units $\hbar=c=k_{B}=1$}.The baryon asymmetry is defined by baryon asymmetry factor (BAF) $\eta$. It is given by \cite{Das2022BG,Davoudiasl2004,Oikonomou2016,Oikonomou2016b,Nozari2018,Bhattacharjee2020}
\begin{equation}
\label{BA2}
\eta=\frac{n_{b}}{s}\simeq\left.-\frac{15g_{b}}{4\pi^{2}g_{*}}\frac{\dot{R}}{M_{*}^{2}T}\right|_{T_{D}},
\end{equation}
where $g_{*}$ is  the number of degrees of freedom for the particle which contributes to the entropy of the Universe, $g_{b}$ is number of intrinsic degrees of freedom of baryons,  $n_{b}$ and $s$  correspond to baryon number density and entropy density, respectively. The entropy density of the Universe is given by $s=2\pi^{2}g_{*}T^{3}/45$. The baryon asymmetry interaction occurs below the critical temperature $T_{D}$. As can be seen from Eq. (\ref{BA2}), BAF is characterised by the time derivative of Ricci scalar. The Ricci scalar is given by
\begin{equation}
\label{BA3}
R=-8\pi G_{N} (\rho-3p)=-8\pi G_{N}(1-3w)\rho
\end{equation}
for the standard cosmological model. It is clear that the Ricci scalar and its derivative vanish in the radiation-dominated era $\omega=1/3$. Therefore, $\eta$ vanishes. However, this result contradicts with the observations \cite{PDG,Planck,Riotto1998,WMAP,Cyburt2004}.
In order to examine the departure from thermal equilibrium of the Universe, $\rho$ and $p$ are defined by $\rho=\rho_{eq}+\delta \rho$  and $p=p_{eq}+\delta p$. Here, $\rho_{eq}$ and $p_{eq}$ are the energy density and pressure at thermal equilibrium, while $\delta \rho$ and $\delta p$ represent the GUP contributions.  Setting $k=0$ and using $\dot{a}^2/a^2=H^2$ with $\ddot{a}/a=\dot{H}+H^{2}$, the Friedmann equations in Eq. (\ref{FriedmanEqExpanded}) are given by
\begin{equation}
\label{BA4}
H^{2}=\frac{8\pi G_{N}\rho}{3}+\frac{2\alpha\pi^{2}G_{N}^{3}}{9}\frac{\rho\left(\rho+3p\right)^{2}}{2\rho+3p}+\mathcal{O}\left(\alpha^{2}\right),
\end{equation}
\begin{equation}
\label{BA5}
H^{2}+\dot{H}=-\frac{4\pi G_{N}\left(\rho+3p\right)}{3}-\frac{2\alpha\pi^{2}G_{N}^{3}\left(\rho+3p\right)^{2}}{9}+\mathcal{O}\left(\alpha^{2}\right).
\end{equation}
Solving these equations, one can obtain $\rho$ and $p$
\begin{equation}
\label{BA5.1}
\rho=\frac{3H^{2}}{8\pi G_{N}}+\frac{3\alpha H^{2}(H^{2}+\dot{H})^{2}}{64\pi\left(H^{2}+2\dot{H}\right)}+\mathcal{O}\left(\alpha^{2}\right)
\end{equation}
\begin{equation}
\label{BA5.2}
p=-\frac{3H^{2}+2\dot{H}}{8\pi G_{N}}-\frac{\alpha(H^{2}+\dot{H})^{2}\left(3H^{2}+4\dot{H}\right)}{64\pi\left(H^{2}+2\dot{H}\right)}+\mathcal{O}\left(\alpha^{2}\right),
\end{equation}
respectively. As long as the departure from thermal equilibrium is sufficiently small, the expansion is mainly governed by the energy density and pressure in thermal equilibrium. Therefore, we can consider $H^{2}=8\pi G \rho_{eq}/3$ and $\dot{H}=-4\pi G_{N} \left(\rho_{eq} + p_{eq}\right)$. Substituting  $p_{eq}=\omega \rho_{eq}$, $H$ and $\dot{H}$ into Eqs. (\ref{BA5.1}) and (\ref{BA5.2}),  we obtain
\begin{equation}
\label{BA6}
\rho=\rho_{eq}-\frac{\alpha\pi G_{N}^{2}\left(1+3\omega\right)^{2}\rho_{eq}^{2}}{12\left(2+3\omega\right)}+\mathcal{O}\left(\alpha^{2}\right),
\end{equation}
\begin{equation}
\label{BA7}
p=\omega\rho_{eq}-\frac{\alpha\pi G_{N}^{2}(1+2\omega)(1+3\omega)^{2}\rho_{eq}^{2}}{12(2+3\omega)}+\mathcal{O}\left(\alpha^{2}\right).
\end{equation}
Substituting Eqs. (\ref{BA6}) and (\ref{BA7}) into Ricci scalar (\ref{BA3}), one finds
\begin{equation}
\label{BA8}
R=\frac{4\pi G_{N}\rho_{eq}\left[6\left(3\omega-1\right)(3\omega+2)-\alpha\pi G^{2}(3\omega+1)^{3}\rho_{eq}\right]}{3(2+3\omega)}
\end{equation}
Using the continuity equation (\ref{continuityEqu}) and the first Friedmann equation in Eq. (\ref{BA8}), the time derivative of Ricci scalar can be obtained as follows:
\begin{equation}
\label{BA9}
\dot{R}=\frac{16\sqrt{6}\pi^{3/2}G_{N}^{3/2}\rho_{eq}^{3/2}(1+\omega)\left[6-\left(1+3\omega\right)\left(9\omega-\alpha\pi G_{N}^{2}\left(1+3\omega\right)^{2}\rho_{eq}\right)\right]}{3(2+3\omega)}.
\end{equation}
For the radiation-dominated era, $\omega=1/3$, $\dot{R}$ is given by
\begin{equation}
\label{BA10}
\dot{R}=\frac{512}{9}\sqrt{\frac{2}{3}}\pi^{5/2}G_{N}^{7/2}\rho_{eq}^{5/2}\alpha.
\end{equation}
Plugging Eq. (\ref{BA10}) into BAF formula in Eq. (\ref{BA2}), one finds
\begin{equation}
\label{BA11}
\eta=-\alpha\frac{32\pi^{3}g_{b}g_{*}^{3/2}}{405\sqrt{5}}\left(\frac{m_{pl}}{M_{*}}\right)^{2}\left(\frac{T_{D}}{m_{pl}}\right)^{9},
\end{equation}
where we use $G_{N}=1/m_{pl}^{2}$ and $\rho_{eq}=\pi g_{*} T^{4}/30$. This result clearly indicates that GUP effects can break thermal equilibrium. In order to constrain the GUP parameter, we set $T_{D}=2\times10^{16}GeV$, $M_{*}=m_{pl}/\sqrt{8\pi}$, $g_{b}\sim1$ and $g_{*}\sim106$ \cite{Kinney2006}. Then, BAF is given by
\begin{equation}
\label{BA12}
\eta=-2.57\times10^{-21}\alpha.
\end{equation}

\begin{table}[ht]
    \caption{Bounds on the GUP parameter $\alpha$ from various observations.}
    \centering
    \renewcommand{\arraystretch}{1.5}
    \begin{adjustbox}{width=\textwidth}
    \begin{tabular}{l c c c}
        \hline\hline
        Observations & $\eta$ & Das et al. \cite{Das2022BG} & In this study \\ [0.5ex]
        \hline
        Particle Data Group \cite{PDG} & $\eta\leq8.6\times10^{-11}$ & $-3.98\times10^{8}\leq\alpha$ & $-3.35\times10^{10}\leq\alpha$ \\
        BBN \cite{PDG} & $3.4\times10^{-10}\leq\eta\leq6.9\times10^{-10}$ & $-3.19\times10^{9}\leq\alpha\leq-1.57\times10^{9}$ & $-2.68\times10^{11}\leq\alpha\leq-1.32\times10^{11}$ \\
        Planck observations \cite{Planck} & $\eta\leq6.2\times10^{-10}$ &$-2.87\times10^{9}\leq\alpha$&  $-2.41\times10^{11}\leq\alpha$  \\
        Deuterium and 3He abundances \cite{Riotto1998} & $-5.9\times10^{-11}\leq\eta\leq9.9\times10^{-11}$ & $-4.58\times10^{8}\leq\alpha\leq2.73\times10^{8}$& $-3.85\times10^{10}\leq\alpha\leq-2.3\times10^{10}$ \\
        Acoustic peaks in CMB (WMAP) \cite{WMAP} & $\eta\leq6.3\times10^{-10}$ & $-2.92\times10^{9}\leq\alpha$ & $-2.45\times10^{11}\leq\alpha$ \\
        Deuterium and Hydrogenium abundances \cite{Cyburt2004} & $-5.9\times10^{-10}\leq\eta\leq6.3\times10^{-10}$ & $-2.92\times10^{9}\leq\alpha\leq-2.73\times10^{9}$ & $-2.45\times10^{11}\leq\alpha\leq-2.3\times10^{11}$ \\
        \hline
    \end{tabular}
    \end{adjustbox}
    \label{table1}
\end{table}
In Table \ref{table1}, we present the GUP bounds obtained from various observations. The bounds obtained in this study differ from those obtained in Das et al. \cite{Das2022BG} by up to two orders of magnitude. Similarly, we find negative bounds for the quadratic GUP parameter. However, this does not mean that only negative bounds exist in the literature. There are also GUP models that yield positive bounds. For instance, GUP models that include higher order or linear term in $\Delta p$ yield positive bounds \cite{Das2022BG,Feng2022}. Our findings  not only confirm  the previous results of Das et al. \cite{Das2022BG}, but also indicate that the modification of the equipartition law due to the GUP  provides fluctuations which break thermal equilibrium in the radiation-dominated era of the Universe. 

\section{Conclusions and discussions}
\label{Conlc}

In this paper, using the GUP-modified equipartition law of energy, we obtained the Friedmann equations from Verlinde's entropic gravity \cite{Verlinde2011,Cai2010}. First, combining HUP with thermodynamical arguments of horizon, we heuristically derived equipartition law of energy. Then, we extended the prescription for the GUP case and obtained the GUP-modified equipartition law of energy. In order to grasp the effects of GUP, we also investigated the classical and quantum gravity limits of the modified equipartition law. We examined the modified Friedmann equations and found a maximum and finite energy density at the beginning of the Universe, which implies a universe without initial singularity. We also found a maximum Hubble parameter at the beginning of Universe, again implying a non-singular universe. Moreover, we computed the Kretschmann curvature scalar for the maximum energy density. We found that the Kretschmann curvature scalar is finite, indicating the absence of an initial singularity. In addition to the positive GUP parameter, our analysis also revealed that a non-singular universe can be described for the negative GUP parameter. We deduced that the negative GUP parameter should be preferable since the initial acceleration requires the condition $\ddot{a}/a>0$. Furthermore, we computed the deceleration parameter. We obtained the minimum scale factor and maximum redshift parameter from the deceleration parameter. These results clearly indicates a non-singular universe. We also obtain the constraints on the equation of state parameter $\omega$ from  $a_{min}$ and $z_{max}$ for negative and positive GUP parameter. Finally, we revisited the investigation of gravitational baryogenesis for our modified Friedmann equations. Similar to previous studies \cite{Das2022BG,Feng2022,Luo2023}, we showed that the GUP-modified equipartition law of energy provides the necessary mechanism for generating baryon asymmetry to break thermal equilibrium in the radiation-dominated era. Similar to the findings of Das et al. \cite{Das2022BG}, we also obtained negative bounds for GUP parameter from observations \cite{PDG,Planck,Riotto1998,WMAP,Cyburt2004}. Another interesting aspect of the GUP-modified Friedmann equations is that they share some similarities with braneworld cosmology \cite{Binetruy2000,Binetruy2000b,Ida2000,Shiromizu2000,Ichiki2002,Bernal2020}. In particular, the first order quantum gravity correction due to the GUP appears as quadratic correction in $\rho$, which resembles the $\rho^{2}$ term in braneworld cosmology.

One of the most remarkable findings of this paper is that GUP-modified Friedmann equations imply a non-singular universe for the negative value of the quadratic GUP parameter. Although a non-singular universe still exists for the positive value of quadratic GUP, it does not satisfy the initial acceleration $\ddot{a}/a>0$. In fact, the negative GUP parameter (\ref{GUP}) does not yield minimal position uncertainty by itself. However, a nonsingular universe emerges due to the dependence of maximum energy density on both $\omega$ and $\alpha$. Usually, non-singular behaviour of a universe occurs  for the positive value of GUP parameter \cite{Awad2014,Salah2017,Okcu2020,Alsabbagh2023,Okcu2024}, especially in the  Friedmann equations that are obtained from modified entropy-area relation. In contrast to Refs. \cite{Awad2014,Salah2017,Okcu2020,Alsabbagh2023,Okcu2024}, a nonsingular behaviour emerges for the negative GUP parameter due to the modifications of equipartition law of energy. This result may provide new insight into the potential role of the negative GUP parameter for curing the initial singularity. 

Although most studies have focused on the effects of the positive GUP parameter, there has been an increasing interest in the investigations of the negative GUP parameter \cite{Zhou2022,Jizba2010,Ong2018,Carr2015,Scardigli2015,Okcu2021}. Jizba et al. \cite{Jizba2010} obtained the negative GUP parameter in the context of the crystal lattice. Zhou et al. \cite{Zhou2022} explored the effects of negative GUP parameter on the black hole phase transition. In Ref. \cite{Ong2018}, Ong showed that there is a Chandrasekhar limit for the negative GUP parameter. Furthermore, he found a finite temperature when a black hole completely evaporates. Carr et al. \cite{Carr2015} investigated sub-Planckian black holes for the negative GUP parameter. In Ref. \cite{Scardigli2015}, Scardigli and Casadio examined the light deflection and perihelion precession for the GUP-modified Schwarzschild metric. They obtained negative GUP bounds from observations. Similar to \cite{Scardigli2015}, we also investigated Shapiro time delay, gravitational redshift and geodetic precession for the GUP-modified Schwarzschild metric \cite{Okcu2021}. We found a negative GUP parameter from the observations as well. As a result, our study reveals a new feature of the negative GUP such as the implication of a non-singular universe, thereby enabling a new perspective on the phenomenology of quantum gravity.

Although the paper has mainly been focused on the derivation of Friedmann equations within the GUP-modified equipartition law of energy and its early universe implications such as the removal of initial singularity and baryogenesis, it is worth nothing that GUP correction may have possible broader consequences for the late time cosmology. In particular, GUP may mimic dark energy since it provides a dynamical dark energy component in addition to the cosmological constant \cite{Luciano2025}. Moreover, the GUP parameter can be constrained through the dark energy equation of state parameter from Planck observations \cite{Planck}, enabling the possible link between late-time cosmology and phenomenology of quantum gravity. On the other hand, GUP may play an important role on the growth of matter perturbation in the early Universe \cite{Luciano2025}. GUP  was also proposed as a possible accounting for dark matter like behaviour at galactic scale \cite{Bosso2022}. A detailed analysis of these issues may be twofold: on the one hand determining the possible consequences of such topics within our setup, on the other investigating the theoretical implications of the GUP for cosmology. In particular, it may be possible to develop new insights for these issues within the framework of GUP-modified equipartition law of energy. The above mentioned issues within our setup will be the subject of future research.

\section*{Acknowledgements}
The author is grateful to the anonymous reviewers for their valuable comments and constructive suggestions, which have significantly improved the clarity and quality of this paper.

\section*{Declaration of competing interest}
The authors declare that they have no known competing financial
interests or personal relationships that could have appeared to influence
the work reported in this paper.

\section*{Data availability}
No data was used for the research described in the article.

\end{document}